\documentclass[12pt]{iopart}
\usepackage{graphicx}
\usepackage{epsfig}

\begin{document}

\title[Beam displacement measurements]{Nano-displacement measurements
using spatially multimode squeezed light}

\author{N. Treps\dag \ddag\footnote[3]{To whom correspondence should be
addressed (nicolas.treps@spectro.jussieu.fr)}, N. Grosse\ddag, W.~P.
Bowen\ddag, M.~T.~L. Hsu\ddag, A. Ma\^\i tre\dag, C. Fabre\dag, H.-A.
Bachor\ddag~and P.~K. Lam\ddag }

\address{\ddag\ Australian Centre for Quantum-Atom Optics, Department
of Physics, Australian National University, Canberra ACT 0200,
Australia.}

\address{\dag\ Laboratoire Kastler Brossel, Universit\'e Pierre et
Marie Curie, case 74, 75252 Paris cedex 05, France }

\begin{abstract}

We demonstrate the possibility of surpassing the quantum noise limit
for simultaneous multi-axis spatial displacement measurements that have
zero mean values.  The requisite resources for these measurements are
squeezed light beams with exotic transverse mode profiles.  We show
that, in principle, lossless combination of these modes can be achieved
using the non-degenerate Gouy phase shift of optical resonators.  When
the combined squeezed beams are measured with quadrant detectors, we
experimentally demonstrate a simultaneous reduction in the transverse
$x$- and $y$- displacement fluctuations of 2.2~dB and 3.1~dB below the
quantum noise limit.

\end{abstract}


\submitto{\JOB}

\maketitle

\section{Introduction}

Optical images are one of the most important means used to convey
information.  Quantum fluctuations are inevitably present in the
measurement performed on each of the pixels used to record the image. 
This results in a deterioration in the quality of the information that can be
extracted from it.  Quantum fluctuations depend on the characteristics
of the light used in the measurement, and specifically designed
non-classical states of light must be used to reduce the quantum noise,
extending the limits of information read-out from these images.  The
study of such quantum fluctuations and correlation in images, and of
their applications, is now a topic of growing attention from the
quantum optics community \cite{LGB, Teich, Lee, kolobov}.

The implementation of a general scheme to reduce quantum noise depends
largely on the specifics of the information to be extracted from the
beam.  For instance, in order to improve the optical resolution (i.e.
reduce the size of the minimum detail distinguishable in the image),
one requires the use of squeezed states of light in a large number of
specific transverse modes \cite{KoloFabre,KoloSoko, LugiatoGrangier} or
N-photon correlated states \cite{Nphoton}.  These states of light are
typically difficult to produce.  However, the problem is greatly
simplified when one possesses {\it a priori} knowledge about the image
to be extracted.  For example, {\it a postiori} information processing
performed on the intensity values measured by each pixel of the
detector can be implemented.  This technique is widely used when
measuring the motion of a microscopic transparent particle \cite{Denk},
macro-molecule \cite{Kamimura, Jelles} and atomic scale surface 
structures \cite{AFM}.  Such measurements can
yield sensitivity in the nanometer regime, beyond the wavelength of the
light used.

In this paper, we will focus on the specific problem of light beam
positioning.  The average transverse intensity of the beam is assumed
to be constant, but its position and orientation in the transverse
plane are unknown.  Such measurements are widely used in many practical
applications, especially in nanotechnology, where accuracy and
reproducibility of positioning at atomic scales are required.  If
coherent light is used, the spatial randomness of photons introduces
quantum fluctuations in the position measurement.  The sensitivity of
positioning is then at the quantum noise limit \cite{AFM}.  It
was proposed theoretically \cite{Fabre} and later demonstrated
experimentally (in one \cite{PRL} and two \cite{Science} dimensions)
that this limit could be circumvented by using light that has spatial
ordering of the photons.  The motivation of this paper is to present an
overview of optical beam positioning, clearly identify the sources of
noise, and then propose ways to improve on these measurements.  This
will enable us to demonstrate the optimisation of the measurement
strategy.  We finally present a detailed account of the experiment
outlined in Ref.~\cite{Science}, and demonstrate that simultaneous 
below quantum noise measurements of the beam centre for two different spatial
axes are possible.

\section{Theory}

\subsection{Signal and Noise in Displacement Measurements}
In many applications, such as atomic force microscopes or photonic
force microscopes \cite{Senden,Rohrbach}, beam positioning is measured
using quadrant detectors.  The sum and difference photocurrents
between the quadrants give information on the relative position of the
beam.  This measurement delivers photocurrents whose mean value is the
positioning {\it signal}, and fluctuations around the mean is {\it
   noise} which limit the accuracy of positioning.

The theoretical treatment of quantum noise arising from 1D position
measurements using split detector was first examined by {\it Fabre et
   al.} \cite{Fabre}.  They showed that the quantum noise could be
reduced using a squeezed vacuum field in a mode with a $\pi$ phase
shift between the halves impinging on each side of the split detector.
Fields with such characteristic phase shifts have henceforth been
termed {\it flipped modes}.  In this article, we provide another
perspective on that result, demonstrating that the quantum noise can
be reduced simultaneously for several orthogonal spatial measurements
performed on the beam.

We use the standard description of the positive frequency part of the
electric field operator \cite{kolobov}, whose transverse dependence is
decomposed over a complete basis of orthogonal modes $\{u_i\}$, within
the paraxial approximation
\begin{equation}
    \hat E^+(\vec\rho,z,t)
    =i\sqrt{\frac{\hbar\omega}{2\varepsilon_0}}\left[\sum_i\hat
      a_i(z,t)u_i(z,\vec\rho)\right]e^{-i\omega(t-z/c)}
\end{equation}
where $z$ is along the propagation direction, $\vec\rho$ is the
two-dimensional transverse co-ordinate, $\hat a_i$ is the annihilation
operator of a photon in mode $u_i$ and the operators satisfy the
commutation relation
\begin{equation}\label{commute}
     [\hat a_i,\hat a_j^\dagger]=\delta_{ij}
\end{equation}
We then consider the slowly varying envelope
$\mathcal{E}(\vec\rho)$ of the field incident on a plane of
longitudinal position $z$ (omitted in the equations), which can be
expressed as
\begin{equation}
    \hat \mathcal{E}^+(\vec\rho)
    =i\sqrt{\frac{\hbar\omega}{2\varepsilon_0}}\sum_i\hat
      a_iu_i(\vec\rho)
\end{equation}

The photon number operator at a position $\vec\rho$ of the transverse plane
(proportional to $(\hat\mathcal{E}^+)^\dagger\hat\mathcal{E}^+$) is
given by
\begin{equation} \label{N}
     \hat\mathcal{N}(\vec\rho)=\sum_{i,j}\hat a_i^\dagger\hat a_j
     u_i^*(\vec\rho)u_j(\vec\rho).
\end{equation}

The general study of measurements performed with several detectors on
the transverse plane will be the subject of a future publication.
Here, we present a specific approach applicable when the difference is
taken between the intensities on two spatial areas of the beam.  To be
more specific we will assume that
the measurement consists of the difference between the left and right
sides of the beam.  This restriction does not affect the generality of
our results, one may simply change the integration domain to adapt them
to any detector shape.  For a split detector, the photon number
difference is given by
\begin{equation}\label{Nmoins}
     \hat N_-=\int_{x<0}\hat\mathcal{N}(\vec\rho)d^2\rho-\int_{x>0} \hat
     \mathcal{N}(\vec\rho)d^2\rho=\sum_{i,j}I(u_i,u_j)\hat
     a_i^\dagger\hat a_j
\end{equation}
where $x$ is the horizontal co-ordinate and $I(u_i,u_j)$ is defined by
\begin{equation}
     I(u_i,u_j)=\int_{x<0}u_i^*(\vec\rho)u_j(\vec\rho)d^2\rho -
     \int_{x>0}u_i^*(\vec\rho)u_j(\vec\rho)d^2\rho
\end{equation}
This equation shows that the pertinent quantities in the calculation
of noise are the overlap integrals between the different transverse
modes.  In order to conveniently evaluate these integrals, we will
choose an adapted transverse basis as proposed in Ref.~\cite{Fabre,
   PRL}.  The first mode in our basis is chosen to match the transverse
electric field profile of the beam under interrogation, which we
denote by $u_0(\vec\rho) = \mathcal{E}(\vec\rho)/||\mathcal{E}||$
(This may not necessarily be a TEM$_{\rm 00}$ mode).  Since the mean
value of the electric field is completely contained in $u_0$, the
subsequent modes in our basis therefore have no coherent amplitude and
only contain fluctuation terms.  We now choose the mode $u_1$ to be
\begin{eqnarray}\label{flipped}
     u_1(\vec\rho) = -u_0(\vec\rho) & \quad \textrm{if} \quad x<0
     \nonumber \\
     u_1(\vec\rho) = u_0(\vec\rho) & \quad \textrm{if} \quad x>0
\end{eqnarray}
where $u_1$ is referred to as the {\it flipped mode} of $u_0$.  As the origin
of the transverse co-ordinate $x=0$ is chosen to be the centre of the
detector, the flipped mode is orthogonal to $u_0$ only in the case when
the beam is centred on the detector, i.e. when the intensities on the
two areas considered are equal.  As we would like to measure very small
displacements to this configuration, we will consider here that the
beam is perfectly centred.  For any given spatial basis,
Eq.~(\ref{flipped}) can be used to generate a corresponding {\it
flipped mode basis}.  In this basis, we have seen that the mean values
of the annihilation and creation operators are zero for all the modes
except $u_{0}$.  Hence only overlap integrals that contain the first
mode, $I(u_i,u_0)$, contribute to Eq.~(\ref{Nmoins}).  Using
Eq.~(\ref{flipped}), overlap integrals are given by
\begin{equation}
     I(u_i,u_0)=\int u_i^*(\vec\rho)u_1(\vec\rho)d^2\rho=\delta_{i,1}
\end{equation}
which means that only $I(u_1,u_0)$ is non-zero.  The noise contribution
from all other modes are zero.  Using the linearised approximation, the
noise in the differential measurement can then be written as
\cite{Fabre,Treps}
\begin{eqnarray}\label{result}
      \langle\delta\hat N_{-}^{2}\rangle &=& N_{\rm tot}\langle(\delta\hat
      a_{1}^{\dagger}+\delta\hat a_{1})^{2}\rangle \nonumber\\
      &=& N_{\rm tot} \langle (\delta \hat X^{+}_{1})^{2} \rangle
\end{eqnarray}
Hence, noise in this measurement arises only from a given quadrature
(namely $\hat X_1^+=\hat a_1+\hat a_1^\dagger$) of
the flipped mode.  It is important to note that this calculation was
performed for a beam centred at the detector.  However it is still
valid for displacements small compared with the beam waist.

For several simultaneous position measurements performed, for instance
with a quadrant detector, one can apply the same method to show that
noise arises from the corresponding flipped mode associated with the
differential measurement.  It is then possible to write an equation
similar to Eq.~(\ref{result}) for each differential measurement with
its corresponding flipped mode.  The orthogonality ensures that the
reduction of measurement noise associated with each mode can be
independently performed.

\subsection{Engineering Spatially Multimode Squeezed Light}

For a given measurement, the precedent analysis gives the requirement
for improving the sensitivity.  Indeed, one has to determine the
flipped modes involved in the measurement.  These modes have then to
be produced in a squeezed vacuum in order to reduce the noise in the
displacement measurement.  As they are orthogonal it is possible, at
least theoretically, to spatially overlap them without any losses, and
then to spatially overlap the resulting beam with the beam we would
like to measure.  Furthermore, the fact that the flipped modes are
orthogonal implies that it is necessary to use as many squeezed vacuum
sources as the number of measurements we would like to improve
simultaneously.

Several strategies can be adopted to produce low intensity squeezed
beams whose transverse profile is a flipped mode.  Shaping the phase of
an optical beam can be easily done with a phase mask.  However, these
devices can be lossy.  As we need only to perform $\pi$ phase shifts
several schemes have been proposed - one involving a half wave plate,
the other a Sagnac interferometer \cite{Delaubert}.  However, other
devices such as the ones used in adaptive optics can be implemented
\cite{courtial}.  Assuming that the flipped mode can be easily
produced without losses (as was the case in our experiment), a
sensible strategy would be to produce a squeezed beam with an optical
parametric amplifier, and then reshape its phase profile.
However, for more complicated schemes, another possibility could be the initial
production of a coherent state in the desired transverse mode followed by
squeezing using for instance a multimode optical parametric
amplifier \cite{ks, kolobov}.

\begin{figure}[!ht]
           \begin{center}
           \includegraphics[width=6cm]{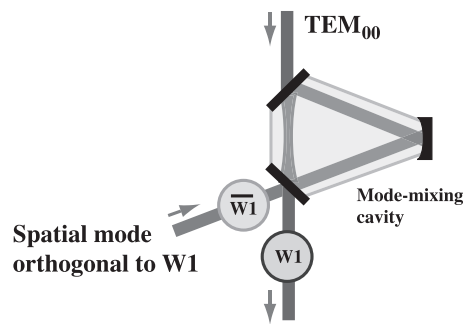}
      \caption{Experimental scheme to spatially overlapping two beams with 
      orthogonal spatial modes using an impedance matched cavity.}
           \label{Cavitymixing}
           \end{center}
\end{figure}

Techniques to effectively spatially overlap the modes are, in general,
dependent on the specific properties of the modes involved, and can be
quite complicated.  However, so long as the modes involved can be
easily produced with wave-plates, any two modes can be overlapped with
high efficiency using the mode dependence of an optical cavity.  One
example of this would be to design a ring optical cavity close to
transverse degeneracy, but far enough away to have a good separation of
the modes.  The cavity can then be tuned to be resonant for the $\rm
TEM_{00}$ mode and far off resonance for all the other modes.

With reference to Fig.  \ref{Cavitymixing}, the mixing of two
transverse modes can be performed in the following way.  We begin with
one field already in its desired mode-shape, and the other in a
TEM$_{00}$ mode transformable to its desired mode shape using the
wave-plate $W1$.  Given that we only perform phase shifts, there exists
a wave-plate $\bar{W1}$ that performs the reverse operation to that
performed by $W1$.  Our $\rm TEM_{00}$ mode is transmitted through an
impedance matched cavity, and the wave-plate $W1$ is placed after the
cavity to produce the desired mode.  The second mode is then reflected
off the empty port of the cavity output mirror.  However, as the mode
will also go through wave-plate $W1$, it is thus necessary to transmit
the mode through wave-plate $\bar{W1}$ prior to the cavity.  The
resulting effect of the wave-plate combination $\bar{W1}W1=I$ ($I$ is
the identity operation) on the second mode, is a preservation of its
initial mode-shape.  In this way, the transmitted and reflected modes
are, in principle, perfectly spatially overlapped, with losses avoided
by an appropriate design for the optical cavity.

Other schemes to spatially overlap the beams can be proposed for
specific cases.  For instance, it often appears that one needs to
mix an odd mode with an even mode.  In this case, a Mach-Zehnder
like interferometer can be utilised, as used in the single photon
community \cite{kk}. In that article, the Mach-Zender was used as a
transverse-mode beam splitter. It is possible to use it the other
way around, as a lossless transverse-mode beam mixer, 
demonstrated in our scheme.

\section{Optimised Displacement Measurement}

\subsection{Beam Imaging}

One of the main applications of precise laser beam positioning is the
measurement of physical effects coupled to the beam displacement. For
instance, in atomic force microscopy, a laser beam is reflected off a
cantilever and its position is recorded by a quadrant detector.  One
can then examine the optimal methods to extract information from the
physical system using that technique.  The previous calculation
demonstrated that, using a coherent beam, the quantum noise in the
measurement depends only on the intensity of the beam.  For a given
intensity, and a given detector shape, a pertinent question is: how
can we obtain the largest signal, and hence the largest signal to
noise ratio?

If we consider only $\rm TEM_{00}$ Gaussian beams, we have shown in a
previous publication \cite{Fabre} that, in the case of a coherent state
impinging on an infinitely broad split detector, the displacement
corresponding to a signal to noise ratio of one is given by
\begin{equation}
       d=\sqrt{\frac{\pi}{8}}\frac{w_0}{\sqrt{N}}
\end{equation}
where $N$ is the total number of photons and $w_0$ the beam waist.
Given this formula, for a fixed laser intensity, it appears that the
relevant quantity is the displacement divided by the beam waist.
Using lenses to increase the beam size, proportionally increases the
relative displacement, and therefore does not change their ratio. The 
consequence of this is that the precision of the measurement will remain the same.
The best configuration to measure a displacement induced by a physical
system is to apply the displacement to the smallest beam possible to
maximise this ratio \cite{AFM}.  The maximum focussing is the
wavelength divided by the numerical aperture of the optical system
\cite{Siegman}. The smallest measurable displacement induced by a
physical system coupled to a coherent laser beam is therefore given by
\begin{equation}\label{dsql}
       d_{SQL}=\sqrt{\frac{\pi}{8}}\frac{\lambda}{2N\!\!A\sqrt{N}}
\end{equation}
where $\lambda$ is the wavelength and $N\!\!A$ the numerical
aperture of the focussing lens.

Let us now consider a detector whose two halves have a finite size $D$
with a gap $\delta$ in between, used to measure a beam displacement.
Changing the beam size with lenses does not affect the precision of
the measurement only in the case of an infinite detector, however here
the losses induced both by the gap and the detector size are
important.  The best signal to noise ratio occurs when minimum
loss is experienced by the beam.  In this case it it therefore
important to optimise the beam waist size to minimise the total losses
induced by the gap and the detector edges.

\subsection{Data Acquisition}

In practice, at low frequencies the quantum noise limit for
displacement measurements is overwhelmed by classical noise arising
from acoustic or seismic vibration sources coupling to the system.
Therefore measurements of the amplitude of periodic displacements
(i.e. displacement modulation) were chosen to be made at high
frequencies $\nu_{\rm mod}>1\rm{MHz}$ where such noise sources were
negligible.  An electronic spectrum analyser (ESA)
can be chosen to extract the spectral power density of the
displacement signal. This corresponds to a measurement of the
oscillation amplitude rather than an absolute displacement.

The refined task is then to measure the square of the amplitude of
the periodic displacement within a small frequency range of
oscillations as a function of time.  An ESA has two parameters
that require specification (RBW and VBW).  The resolution
bandwidth (RBW in units of kHz) determines the small range of
frequencies included in the measurement around the central
frequency $\Omega_{\rm mod}$.  The band-passed signal is then
demodulated at $\Omega_{\rm mod}$ and its amplitude detected. The
measurement time $T$ is given by the inverse of the RBW setting.
$N$ photons are collected in this time, setting the level of noise
power measured on the ESA. In the absence of displacement
modulations, this level corresponds to the quantum noise, and as such
is used as a reference and calibration for further displacement
modulation measurements.

The response of an ESA is non-trivial for the combination of Gaussian
noise and a signal \cite{IEEE}.  For a displacement modulation signal
$d_{\rm mod}$ and noise $d_{\rm noise}$ of the same order $(d_{\rm
mod}\approx d_{\rm noise})$, the response of the ESA in units of
displacement squared may be approximated as $(d_{\rm total}^{2}=d_{\rm
mod}^{2}+d_{\rm noise}^{2})$.  The calibration from spectral noise
power to displacement is determined as follows - for a laser beam of
power $P$, frequency $\nu$, waist size $\omega_{0}$, and an ESA
setting of $\rm{RBW}$, the number of photons collected per measurement
is given by $N=P/h\nu {\rm RBW}$, which determines $d_{\rm SQL}$
from Eq. (\ref{dsql}).

In order to set the calibration constant, a trace from the ESA was
obtained where $d_{\rm mod}$ was kept to zero, hence giving the noise
power level corresponding only to the quantum noise limit
contribution.  Another ESA trace of $d_{\rm total}$ where $d_{\rm
mod}$ was non-zero was then obtained.  The smallest identifiable
displacement ($d_{\rm min}$) can be defined in terms of the
statistical confidence that a displacement modulation has been
observed.  One complication of using an ESA to identify displacement modulations
is that the video bandwidth (VBW in units of Hz) operation of an ESA
sets the averaging length over several $\Omega_{mod}$ oscillation
cycles of $d_{total}$, and thus changes the time response of our
displacement measurement.  However, for practical applications where we
require a time resolution equal to the one set by the resolution
bandwidth, we simply choose the video bandwidth equal to the
resolution bandwidth.

In summary, an efficient quantum noise limited displacement modulation
measurement can be achieved by carefully focusing the beam on the
sample and detector, and choosing appropriate spectrum analyser
settings.  After optimisation of these parameters, there remain only
two ways to improve the quality of the measurement - increasing the
mean power or using spatial squeezing.  It is the latter solution that
we implement and describe in the following section.

\section{2-D Spatial Squeezing - Experimental Scheme and Results}

\subsection{Quadrant Detection}

\begin{figure}[!ht]
       \begin{center}
       \includegraphics[width=8.5cm]{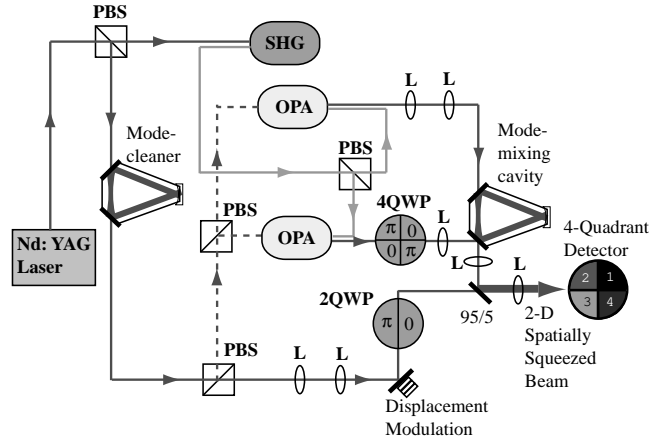}
       \caption{Schematic of experimental setup for the production of
       a 2-dimensional spatially squeezed beam and its detection
       scheme.  PBS: polarising beam-splitter, SHG: second-harmonic
       generator, OPA: optical parametric amplifier, 4QWP:
       4-quadrant wave-plate, 2QWP: 2-quadrant wave-plate, L:
       imaging lens, 95/5: Beam-splitter with 95~\% reflection and 5~\%
       transmission.}
       \label{setup}
       \end{center}
\end{figure}

Our quadrant detector, composed of a $2\times2$ array of separate square detectors
of side 0.5~mm and adjacent separation of 25~$\mu$m (Epitaxx ETX
505Q), allows us to perform 4 independent measurements (see Fig.
\ref{setup})
\begin{eqnarray}
       I_a & = & I_1+I_2+I_3+I_4 \nonumber \\
       I_b & = & (I_1+I_4)-(I_2+I_3) \nonumber \\
       I_c & = & (I_1+I_2)-(I_3+I_4) \nonumber \\
       I_d & = & (I_1+I_3)-(I_2+I_4)
\end{eqnarray}
The first measurement, $I_a$ corresponds to the sum of the four
quadrants and gives the total intensity. $I_b$ and $I_c$
correspond respectively to horizontal and vertical positioning.
Finally, $I_d$, which is the difference between the two diagonal,
can be interpreted, to the first order, as an orientation
measurement. Indeed, if the beam is for instance elliptical, $I_d$
is zero only if the axis of the ellipse is parallel to the
quadrants. To each of these measurements corresponds a flipped
mode defined by Eq. (\ref{flipped}). In the case of a $\rm
TEM_{00}$ input beam, the three flipped modes are orthogonal, and
it is then possible to improve all 3 differential measurements
simultaneously. We will show here how we could improve any two of
them simultaneously, in a configuration easily scalable to the
three measurements.
Experimentally, the challenges in this process
are to efficiently produce the correct spatial modes, to generate a
pair of squeezed fields, and to combine these fields with the coherent
horizontally flipped mode with maximum efficiency.  Our experimental
setup, which overcomes these challenges, is shown in Fig.~\ref{setup}.

\subsection{Generation of Squeezed Light}

In our experiment a 1.5~W field at 1064~nm was generated by a
non-planar ring-oscillator Nd:YAG laser.  This field was split into
two roughly equal parts.  One part was used to pump a second harmonic
generator (SHG), and the other was ultimately used to seed a pair of
optical parametric amplifiers (OPAs) each generating a squeezed field,
and to provide the coherent horizontally flipped field.  The SHG
was constructed from a hemi-lithic MgO:$\rm LiNbO_{3}$ crystal with
curved surface AR coated at 1064~nm and 532~nm, and an input/output
coupling mirror with 92~\% reflectivity for 1064~nm light and $\sim
5$~\% reflectivity for 532~nm light.  An electro-optical modulation
was applied to the MgO:$\rm LiNbO_{3}$ crystal, resulting in a phase
modulation on the intra-cavity field.  An error signal for the length
of the cavity was extracted by detecting the light transmitted through
the cavity and mixing down at the modulation frequency.  The SHG was
non-critically phase matched by holding the temperature of the
MgO:$\rm LiNbO_{3}$ crystal constant at $\sim 107^{\circ}C$.  At
this temperature, a conversion efficiency of $\sim 55$~\% was
achieved.  This resulted in 350~mW of optical power at 532~nm, which
was used to pump our OPAs.

The second 1064~nm beam was passed through a mode cleaning cavity.
This cavity had a line-width of 2~MHz and so acted to filter out noise
on the laser above that frequency.  Consequently, the output field
was quantum noise limited at frequencies above $\sim$~6~MHz.  It was
then used to seed both of our OPAs and to provide the horizontally
flipped coherent field.  The OPAs were of identical design to the SHG,
but with higher output coupler reflectivity at 1064~nm (96~\%).  As
with the SHG, the length of each OPA cavity was controlled using an
error signal generated through an intra-cavity phase modulation.  Each
OPA either amplified or de-amplified its seed field, depending on the
relative phase between the pump and seed.  The SHG phase modulation
resulted in a phase modulation on the OPA pump fields, and
consequently a modulation of the OPA amplification.  For each OPA,
an error signal could be extracted from this modulation, locking the
OPA to de-amplification.  In this regime the OPAs each produced a
$150~\mu$W amplitude squeezed field.  These squeezed fields were
observed to be 4~dB and 3~dB below the quantum noise limit,
respectively.

\subsection{Manipulation of Transverse Mode-Shapes}

At this stage, we had available all of the resources required to
generate a 2D spatially squeezed beam, all that remained was to
appropriately modify each beam's transverse mode-shape, to
generate a spatial modulation, and to spatially overlap the beams.
In our experiment, for practical reasons, instead of using a
TEM$_{\rm 00}$ laser we chose to have the mean field in a horizontally
flipped mode TEM$_{\rm f00}$. Hence, the flipped mode for horizontal
positioning is simply a TEM$_{\rm f00}$ mode and the one for vertical
positioning is a doubly flipped mode TEM$_{\rm f0f0}$. To summarise,
we required the bright coherent field to have a horizontally phase
flipped mode-shape $\rm TEM_{\rm f00}$, and the dim squeezed fields to
have $\rm TEM_{\rm 00}$, and both horizontally and vertically flipped
$\rm TEM_{\rm f0f0}$ mode-shapes, respectively.
The required mode-shape modification could be
achieved by phase flipping some quadrants of each field. This
phase flipping was achieved using birefringent optical half
wave-plates which were cut and assembled at appropriate $\pi/2$
angles (CSIRO Lindfield Australia).  A 2-quadrant wave-plate was
used on the local oscillator beam so that its mode-shape was
horizontally phase flipped, while a 4-quadrant wave-plate was used
on one squeezed beam so that its mode-shape was both horizontally
and vertically flipped. The other squeezed beam required no
modification and was maintained as a $\rm TEM_{00}$ Gaussian mode.

\begin{figure}[!ht]
           \begin{center}
           \includegraphics[width=8.5cm]{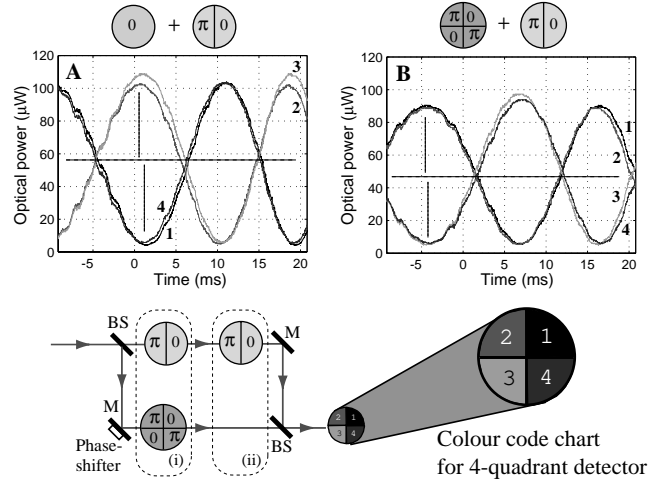}
       \caption{Plot A: interference of $\rm TEM_{00}$ mode with $\rm
       TEM_{f00}$ mode. Plot B: interference of $\rm TEM_{f0f0}$ mode
       with $\rm TEM_{f00}$ mode.  The vertical lines indicate the times when 
       constructive and destructive interference occurs.
       The Mach-Zehnder setup is shown where wave-plate combination
       (i) interrogates the interference between the $\rm TEM_{f0f0}$ and
       $\rm TEM_{f00}$ modes and (ii) interrogates the interference between
       the $\rm TEM_{00}$ and $\rm TEM_{f00}$ modes. The numbers (1,2,3,4)
       on the plots indicate the measurement results corresponding to the
       quadrants of the detector. BS: beam-splitter, M: mirror}
           \label{interference}
           \end{center}
\end{figure}

In order to gauge the quality of the phase flipped wave-plates
fabricated, we interfered the flipped modes with either another
flipped mode or the $\rm TEM_{00}$ mode (see Fig.
\ref{interference}).  This was performed using a Mach-Zehnder
interferometer setup where one arm of the interferometer had a $\rm
TEM_{f00}$ wave-plate and the other arm had either a $\rm TEM_{f0f0}$
wave-plate or no wave-plate.  The input modes at the interferometer
were of equal optical power and by scanning the relative phase between
the beams in the two arms, interference fringes on each quadrant of
the 4-quadrant detector (labelled $1$, $2$, $3$ and $4$ here as
defined in Fig.~\ref{interference}) were observed.  We quantified the
quality of the mode-matching via the interference fringe visibility
\begin{equation}
V = \frac{I_{\rm max} - I_{\rm min}}{I_{\rm max} + I_{\rm min}}
\end{equation}
where $V=1$ indicates perfect mode-matching and $V=0$ indicates no
mode-matching.

When a $\rm TEM_{00}$ mode was interfered with a horizontally flipped
mode (as shown in Fig.~\ref{interference}~A) the interference observed
on the half-detector defined by quadrants 2 and 3 was $\pi$ out of
phase with that on the half-detector defined by quadrants 1 and 4, as
expected since the phase of the $\rm TEM_{f00}$ field was flipped by
$\pi$ between the halves.  The measured visibility of the interference
was $V=0.90$ and the phase mismatch between the constructive and
destructive interference was $7^{\circ}$.  As expected, when a $\rm
TEM_{f0f0}$ mode was interfered with the $\rm TEM_{f00}$ mode (as
shown in Fig.~\ref{interference}~B) the interference observed on the
half-detector defined by quadrants 1 and 2 was $\pi$ out of phase with
that on the half-detector defined by quadrants 3 and 4.  In this case,
the visibility of the interference was $V=0.88$ and phase mismatch was
$5^{\circ}$.  The effective efficiency of our mode-shape modification
procedure is given by $V^{2}$, so that the efficiency for both
wave-plate designs was approximately 80\%.
%

\subsection{Overlapping the Spatial Modes}

To obtain a single 2D spatially squeezed beam, the three modes
produced above were spatially overlapped.  Since squeezing is
highly sensitive to inefficiencies, it was critical that the
efficiency of the process overlapping the squeezed modes was high.
An element that accurately distinguished between different spatial
modes was required to achieve this high efficiency.  In our case
we chose a specially designed optical cavity in a ring
configuration as this element. Cavities constructed using
spherical mirrors decompose incident light fields into the
TEM$_{\rm mn}$ modes.  The Gouy phase shift between the TEM$_{\rm
mn}$ modes typically ensures that for a given cavity length some
TEM$_{\rm mn}$ modes will be resonant, and others entirely
non-resonant.  In our case, we designed the cavity to be impedance
matched, and have relatively low finesse ($\sim 31$).  The input
and output couplers had 95~\% reflectivity and the third mirror
was HR coated.  In this regime losses in the cavity were small
compared to the input and output coupling, and resonant incident
modes were transmitted with high efficiency.  With a cavity length
of 200~mm controlled such that the TEM$_{\rm 00}$ mode was
resonant, our TEM$_{\rm 00}$ squeezed beam was transmitted with $>
95$~\% efficiency and a line-width of 25~MHz.  We were interested
in squeezing at sideband frequencies of $\sim 5$~MHz.  This is
well below the line-width of the cavity, ensuring efficient
transmission of our squeezing.

Let us now consider the TEM$_{\rm f0f0}$ squeezed mode interacting
with the cavity.  Since the cavity was in a ring configuration,
reflection of this mode from the cavity output coupling mirror with
appropriate imaging lenses achieved the desired spatial overlap with
the transmitted TEM$_{\rm 00}$ mode (see Fig.~\ref{Cavitymixing}).  Of
course, for this reflection to be efficient the TEM$_{\rm f0f0}$ had
to be far way from the cavity resonance.  The resonance condition of
the TEM$_{\rm f0f0}$ mode for the ring-cavity can be examined by first
expanding it in the $\rm TEM_{\rm mn}$ basis
\begin{eqnarray}
{\rm TEM_{f0f0}} = \sum_{i=0}^{\infty} \sum_{j=0}^{\infty} c_{mn} {\rm
TEM_{mn}} \\c_{mn} = \frac{(-1)^{i+j}(2i)!(2j)!}{\pi (i!)  (j!)
2^{i+j-1} \sqrt{(2i+1)!  (2j+1)!}}
\end{eqnarray}
where $m=2i+1$ and $n=2j+1$.  Notice that there is no contribution to
the TEM$_{\rm f0f0}$ mode from the TEM$_{\rm 00}$ mode.  Thus, these
two modes are orthogonal, and there is no fundamental limitation on
the efficiency with which they can be spatially overlapped.  A number
of cavity designs exist for which this overlap, in principle,
approaches 100~\% efficiency.  The crucial factor is that when the
TEM$_{\rm 00}$ mode is on resonance, all higher order TEM$_{\rm mn}$
mode with significantly large occupation $|c_{mn}|^{2}$ are far off
resonance.  In the limit of high cavity finesse this condition is
naturally satisfied.

In our case, however, a relatively low cavity finesse was chosen to
ensure the efficient transmission of squeezing at 5~MHz on the
TEM$_{\rm 00}$ squeezed mode.  In this case, a careful analysis of the
resonance condition of each of the significant higher order TEM$_{\rm
mn}$ modes is required.  For each mode, this resonance condition is
dictated by the Gouy phase shift experienced by the modes, which can
be optimised for a given cavity length through an appropriate choice
of the radii of curvature of the cavity mirrors.  For our cavity the
input/output coupling mirrors were chosen to be flat and the third
mirror to have a radius of curvature of 250~mm.  With the cavity
length controlled so that the TEM$_{\rm 00}$ mode was resonant, all
higher order modes with $n+m<18$ were efficiently reflected.  The
contribution to our TEM$_{\rm f0f0}$ mode from TEM modes with $n+m>18$
was insignificant with $|c_{mn}|^{2}<0.0003$ in all cases.  For our
cavity configuration, therefore, the TEM$_{\rm f0f0}$ mode was
efficiently reflected.  We directly observed an efficiency $> 94$\%
for this process.

\begin{figure}[!ht]
           \begin{center}
           \includegraphics[width=7cm]{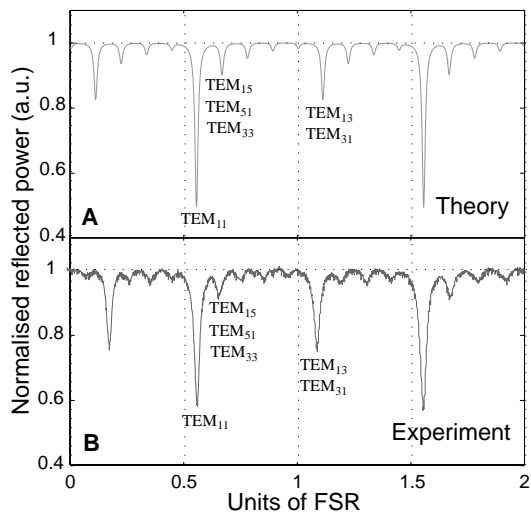}
       \caption{Theoretical (A) and experimental (B) intensity plots for
       $\rm TEM_{f0f0}$ reflection from the ring-cavity.  For the
       theoretical reflection plot, the reflected intensity was computed
       for the first $m=n=29$ $\rm TEM_{mn}$ modes with cavity parameters
       equal to the cavity used in the experiment.}
           \label{cavity}
           \end{center}
\end{figure}

Fig.~\ref{cavity} shows a comparison of the predicted and
experimentally observed cavity reflection for a TEM$_{\rm f0f0}$ mode
as a function of the cavity resonance condition.  The cavity resonance
condition was scanned by ramping the cavity length using a
piezo-electric actuator (PZT) attached to the curved cavity mirror.
When a TEM$_{\rm mn}$ mode was resonant, it was almost entirely
transmitted, and the reflected power from the cavity dropped in
consequence.  The magnitude of the intensity drop was directly
proportional to $|c_{mn}|^{2}$.  Fig.~\ref{cavity}, therefore,
illustrates the significance of each of the lower-order TEM$_{\rm mn}$
modes constituting the TEM$_{\rm f0f0}$ mode.  The highest
contribution came from the TEM$_{\rm 11}$ mode, and the significance
of the modes dropped rapidly as $m+n$ increased.  A comparison of
Fig.~\ref{cavity}~A and B shows a good agreement between the theory
and experiment.  The experimental peaks were slightly broader than the
prediction which can be attributed to the intra-cavity loss.

Transmitting the TEM$_{\rm 00}$ squeezed beam through our cavity, and
simultaneously reflecting the TEM$_{\rm f0f0}$ squeezed beam from the
cavity mirror spatially overlapped the two fields.  All that remained
to generate a 2D spatially squeezed beam was to spatially overlap this
combined beam with our bright coherent TEM$_{\rm f00}$ mode.  Since the
second order statistical moments of coherent fields are insensitive to
attenuation, this overlap could be achieved by utilising a 95/5
beam splitter.  The combined squeezed field, and the coherent flipped
field were imaged onto the two input ports of the beam splitter.  The
combined squeezed field was reflected with 95~\% efficiency, and the
flipped coherent field was transmitted with 5~\% efficiency.   the
resulting output field was 2D spatially squeezed.

\subsection{2D Spatial Squeezing Measurements}

\begin{figure}[!ht]
           \begin{center}
           \includegraphics[width=7.0cm]{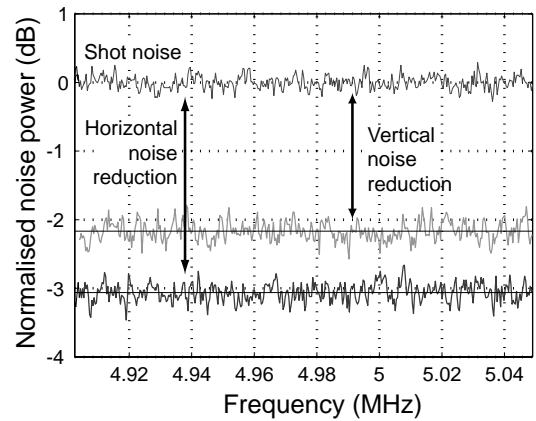}
       \caption{The plot at 0dB (black) is the quantum noise level.  The
       plots at -2.2dB (green) and -3.1dB (red) are the respective
       noise levels for the vertical and horizontal axes of the
       2-dimensional spatially squeezed beam.  Spectrum analyser
       settings: 150~kHz span at 4.976~MHz.  RBW=100~kHz, VBW=100~Hz,
       averaged over 10 traces.}
           \label{result1}
           \end{center}
\end{figure}

The level of squeezing exhibited by our spatially squeezed beam was
examined by imaging the mode onto a quadrant detector.  The horizontal
beam displacement fluctuation was obtained by taking the difference
between the photocurrents originating from the left and right halves
of the quadrant detector, and the vertical beam displacement
fluctuation was obtained from the difference of the photocurrents from
the top and bottom halves of the detector.  These photocurrents were
then analysed in a spectrum analyser which performed power spectral
density measurements.  Without squeezing, these measurements were
limited to the quantum noise level as shown by the black trace in
Fig.~\ref{result1}.  Introducing the squeezed fields, we observed
simultaneous squeezing of both the horizontal and vertical
displacement fluctuations as shown by the colour (grey) traces in
Fig.~\ref{result1}.  The optimum simultaneous squeezing for the
vertical and horizontal displacement fluctuations were 2.0$\pm$0.1~dB
and 3.05$\pm$0.1~dB, respectively.  This corresponded to a reduction
in the displacement fluctuations of the vertical and horizontal axes
by 37~\% and 50~\%, respectively, with respect to the quantum noise
level.
%

\subsection{Implementation and Measurement of a Displacement Modulation}

With a reduction in the noise levels of both the horizontal and
vertical axes to below the quantum noise limit, we now examine the
effects of an externally applied displacement modulation on the
spatially squeezed beam.  Spatial squeezing should improve the
resolution with which this modulation can be analysed.

The external displacement modulation was applied to the $\rm
TEM_{f00}$ local oscillator beam by reflecting the beam from a
$45^{\circ}$ angled mirror mounted on a PZT modulated at 4.976~MHz.
The PZT consisted of a stack of piezo-electric crystals.  When a
voltage was applied across the PZT, the PZT not only
contracted/expanded in the principal axis but also tilted at an angle
away from the principal axis.  The resulting transverse modulation
from the PZT was a diagonal displacement modulation, with horizontal
and vertical components.

\begin{figure}[!ht]
           \begin{center}
           \includegraphics[width=6.5cm]{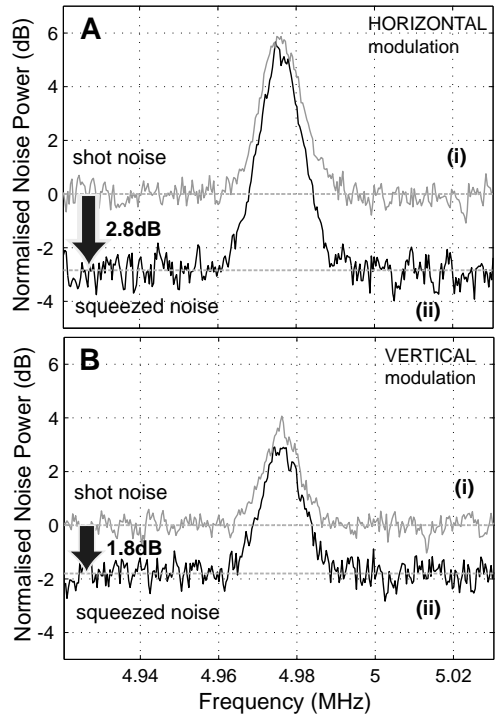}
       \caption{Curves (i) in plots A and B (yellow) are the measured
       displacement modulation for a coherent beam in the horizontal
       and vertical axes, respectively.  Curves (ii) in plots A and B
       (black) are the measured displacement modulation for a
       2-dimensional spatially squeezed beam in the horizontal and
       vertical axes, respectively.  Spectrum analyser settings :
       150~kHz span at 4.976~MHz, RBW=10~kHz, VBW=100~Hz, averaged over
       20 traces.}
           \label{result2}
           \end{center}
\end{figure}

In order to demonstrate an improvement in the resolution measurement
of our 2-dimensional spatially squeezed beam, we compared the results
of a displacement modulation measurement between a coherent beam and
our spatially squeezed beam.  The results are shown in Fig.
\ref{result2}.  The squeezing levels obtained for the horizontal and
vertical axes were 2.84~dB and 1.80~dB, respectively.  The broad peak
corresponded to the measured displacement modulation signal and we
measured our signal to noise ratio (SNR) from the maximum of the
signal peak to the noise floor.  For the quantum noise limited coherent
beam (curve (i)), the measured SNR for the horizontal and vertical
axes were 2.8 and 1.4.  For the 2-dimensional spatially squeezed beam
(curve (ii)), the measured SNR for the horizontal and vertical axes
were 5.2 and 1.9.  The improvement in the measurement of the
displacement modulation signal obtained from the spatially squeezed
beam over the coherent beam is given by the ratio of $\rm
SNR_{sqz}^{H(V)}/SNR_{coherent}^{H(V)}$.  The respective ratios
obtained for the horizontal and vertical axes were 1.9 and 1.4.

\begin{figure}[!ht]
           \begin{center}
           \includegraphics[width=7.5cm]{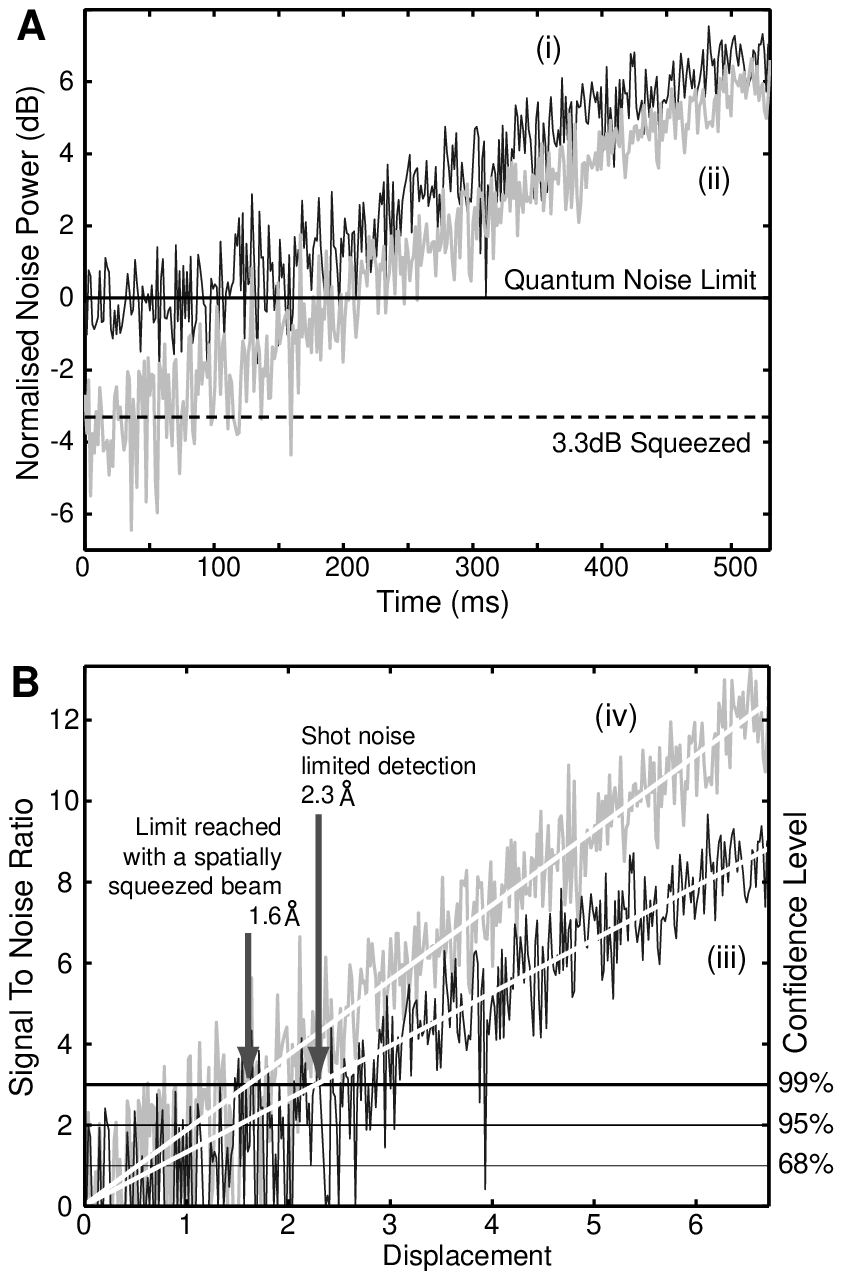}
       \caption{Plot A shows the results of a horizontal displacement
       measurement signal ramped up in time, with (i) obtained from a coherent beam
       and (ii) from a spatially squeezed beam.  In plot B, the
       signal-to-noise improvement (left vertical axis) is plotted
       against the inferred displacement.  Traces (iii) and (iv) show
       the results from data traces (i) and (ii), respectively.
       Spectrum analyser settings: RBW = VBW = 1~kHz, averaged over 20
       traces each with detection time $\triangle t = 1$~$ms$ per data point.}
           \label{noise}
           \end{center}
\end{figure}

We also examined the effects of ramping the displacement modulation
amplitude over time.  To determine the displacement modulation
amplitude at an analysis frequency $\Omega$, the photocurrents from
the quadrant detector were analysed using a spectrum analyser.  The
spectrum analyser demodulated the modulation signal and then performed
a measurement of the power spectral density.  The displayed
information was the sum of the squares of the quantum noise without
modulation $d_{\rm noise}(\Omega)$ and the small transverse beam position
modulation $d_{\rm mod}(\Omega)$ - i.e. $d_{\rm noise}(\Omega)^{2} +
d_{\rm mod}(\Omega)^{2}$ (see
Fig. \ref{noise}A).  Curve (i) is the result of the measurement
performed using a coherent state and thus is the best classical
measurement.  Curve (ii) was obtained by performing the same
measurement, but with a spatially squeezed beam instead.

The data obtained in Fig.~\ref{noise}A was then normalised to the
respective noise levels for the coherent and the spatially squeezed
beams, shown in Fig.~\ref{noise}B. The vertical axis is the
difference between the measured signal, and noise without displacement.
For the spatially squeezed beam the average of the signal trace
crosses the threshold of confidence for a smaller modulation amplitude
than for the coherent beam.  The corresponding oscillation amplitude
for the spatially squeezed and coherent beams were found to be
1.6~\AA~and 2.3~\AA~at the 99~\% confidence level.  Thus the
improvement for the spatially squeezed beam was a factor of 1.5 over
the quantum noise limited coherent beam.  Since both traces increased
linearly with the modulation amplitude, this improvement factor was
independent of the choice of confidence levels.

For the sake of brevity, we have only included the results for the
horizontal axis.  It should be realised that an equivalent improvement
for two simultaneous measurements performed in two orthogonal
quadratures on the spatially squeezed beam has been achieved.

\subsection{Simultaneous 2D Displacement Measurement}

\begin{figure}[!ht]
           \begin{center}
           \includegraphics[width=8.5cm]{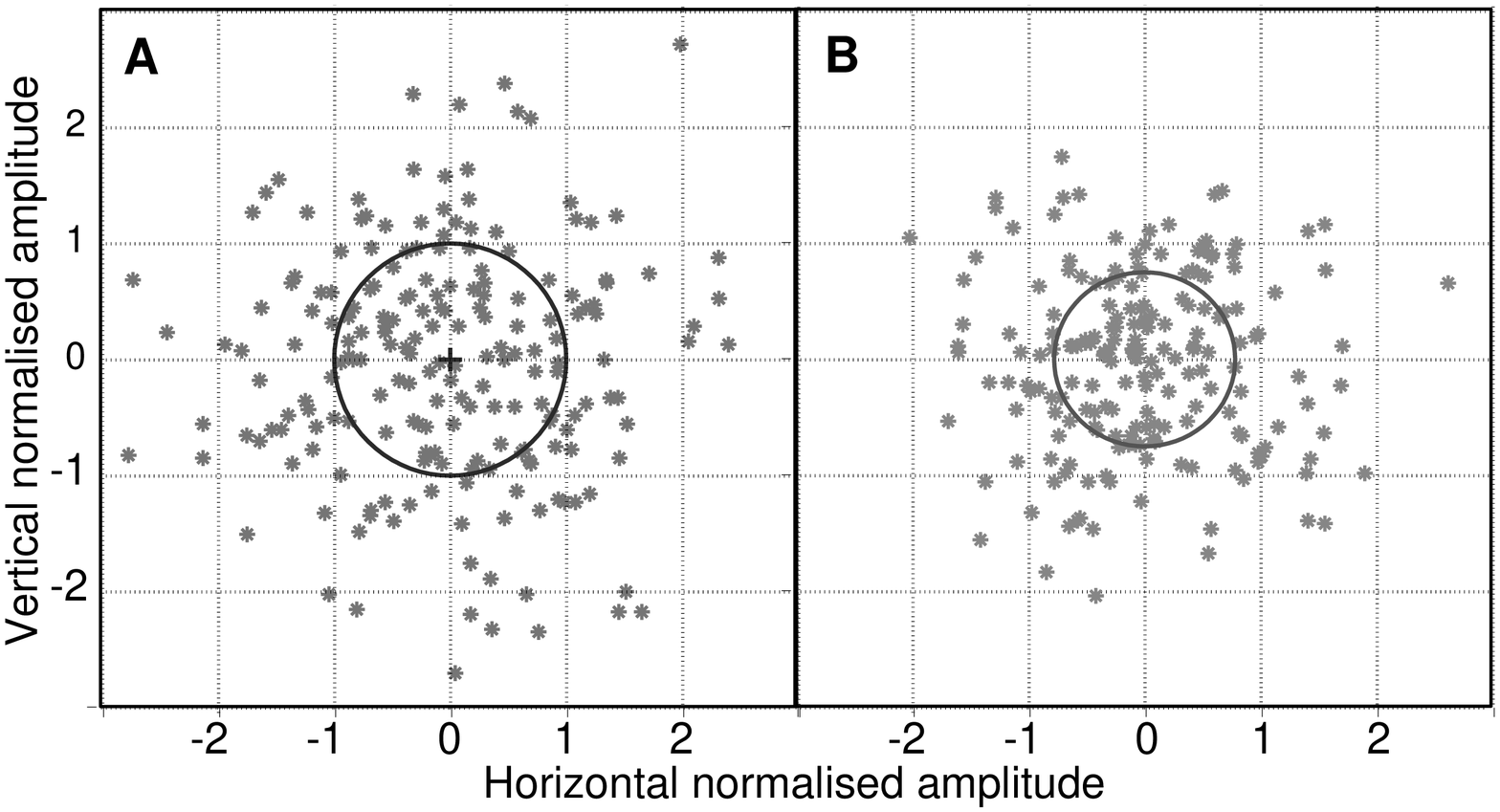}
       \caption{The quantum noise data points (blue) are shown in plot A
       and the squeezing points (red) are shown in plot B.  The quantum
       noise mean value is indicated by the blue $+$ sign at the centre
       of plot A.  The blue and red circles indicate the standard
       deviations for the quantum noise and squeezing data, respectively.
       Re-scaling was implemented such that the quantum noise standard
       deviation equalled one.  The two spectrum analysers had settings
       of 0~Hz span at 4.976~MHz, RBW = VBW = 1~kHz, averaged over 6
       traces.}
           \label{correlation}
           \end{center}
\end{figure}

An intuitive visualisation of our 2D spatial squeezed beam was
obtained by analysing the vertical and horizontal normalised amplitude
quadratures on a correlation diagram.  To do this, simultaneous
measurements of the horizontal and vertical beam displacements were
required.  We achieved these measurements using a pair of
simultaneously triggered spectrum analysers with identical settings.
The results for the two axes could then be plotted against each other
in conventional correlation diagrams.  Fig.~\ref{correlation} shows
the resulting correlation diagrams for a quantum noise limited beam
(Fig.~\ref{correlation}~A) and our spatially squeezed beam
(Fig.~\ref{correlation}~B).  Each point in these diagrams can be
interpreted as an instantaneous 2D measurement of the fluctuations of
the displacement modulation.  The standard deviation as a function of
modulation angle could be calculated for each correlation diagram, and
is displayed by the ellipses in the figure.  We can directly see that
the average displacement modulation fluctuation for all angles is
smaller with our spatially squeezed beam than with a coherent beam.
Specifically, for the horizontal and vertical axes, the average
fluctuation is reduced by a factor of 0.78 and 0.75, respectively.
The square of these values give the fluctuation variances of 2.2~dB and
2.5~dB for the horizontal and vertical axes, respectively.  The
ellipses in Fig.~\ref{correlation} also allow an analysis of any
correlation present between the horizontal and vertical displacement
fluctuations.  Since in both cases the major and minor ellipse axes
correspond well to the horizontal and vertical displacement axes, it
is clear that no cross-axis correlation exists.

\subsection{Extension to Diagonal Axis Measurement}

\begin{figure}[!ht]
           \begin{center}
           \includegraphics[width=6.5cm]{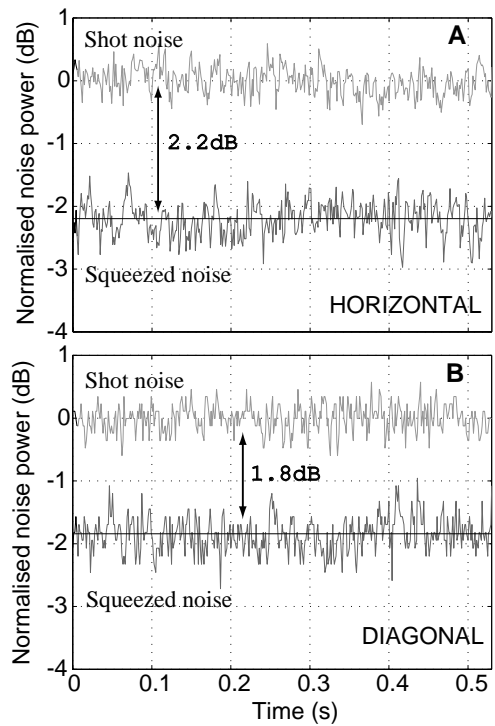}
       \caption{Horizontal (A) and diagonal (B) axes squeezing spectra.
       Spectrum analyser settings: 0~Hz span at 4.976~MHz, RBW=100~kHz,
       VBW=100~Hz, no averaging.}
           \label{result3}
           \end{center}
\end{figure}

As discussed in the theory section, simultaneous enhancement of any
two orthogonal spatial properties of a beam is possible given access
to sufficient squeezing resources, and the ability to choose and
combine appropriate spatial modes.  In this section, we demonstrate
that principle by simultaneously generating horizontal and diagonal
spatial squeezing, contrasting the horizontal/vertical spatial
squeezing discussed throughout the majority of this paper.  Our bright
coherent beam and TEM$_{\rm 00}$ squeezed beam were maintained in their
original modes, and the 4-quadrant wave-plate in the TEM$_{\rm f0f0}$
squeezed beam was replaced by a 2-quadrant wave-plate, rotated by
$90^{\circ}$ relative to the wave-plate for the coherent beam.  This
transformed the TEM$_{\rm f0f0}$ squeezed beam to a vertically flipped
$\rm TEM_{0f0}$ mode.  So that, in this case, the beams involved were
a bright coherent horizontally flipped $\rm TEM_{f00}$ beam, a
vertically flipped $\rm TEM_{0f0}$ squeezed beam and a $\rm TEM_{00}$
squeezed beam.  To measure the diagonal axis modulation displacement,
the photocurrents from diagonal detector quadrants were summed, and
the difference of the two resulting photon currents were analysed in a
spectrum analyser.  The resulting horizontal and diagonal displacement
modulation spectra are shown in Fig.~\ref{result3}.  As can be seen
from the figure, the horizontal and diagonal axes were simultaneously
squeezed by 2.2~dB and 1.8~dB, respectively.  This corresponded to
displacement fluctuation reductions with respect to quantum noise of
40~\% and 34~\%, for the horizontal and diagonal axes respectively.
%

\subsection{Efficient One Dimensional Spatial Squeezing}

In our original one dimensional (1D) spatial squeezing work reported
in Ref.  \cite{PRL}, the spatially squeezed beam was produced by
overlapping a squeezed $\rm TEM_{00}$ beam and a bright coherent $\rm
TEM_{f00}$ beam using a 95/5 beamsplitter.  The choice of 95~\%
reflectivity was motivated by the requirement to minimise loss in the
$\rm TEM_{00}$ squeezed beam, while ensuring a significant fraction of
the coherent beam was transmitted.  However, this resulted in an
attenuation of the displacement signal encoded on the coherent beam,
so that the achievable signal-to-noise ratio was lower than would be
possible if the spatial overlap was achieved efficiently.  We repeated
that original experiment, using our cavity to efficiently spatially
overlap the two beams.  This resulted in a 1D spatially squeezed beam
with minimal losses for both the squeezed and coherent beams.  The
measured squeezing was 2.5~dB, which was also an improvement on the
value of 2.34~dB reported in Ref.~\cite{PRL}.

\section{Conclusion}
In this paper we have investigated, both theoretically and
experimentally, the optimum displacement measurements that can be
performed using a detector array.  We demonstrated that, splitting the
array into two arbitrary areas with equal optical power on each,
enhancement of the difference measurement between the halves could be
achieved using a $\pi$ phase-flipped squeezed mode.  The location of
the $\pi$ phase-flip was dictated by the detector areas under
investigation.  We then showed that an arbitrary number of such
measurements could be enhanced simultaneously using multiple squeezed
beams, so long as the squeezed beams had orthogonal spatial modes.

In our experiment, we were interested in enhancing measurements
performed using an array detector with only four pixels, or in other
words, a quadrant detector.  Such measurements can be interpreted as
displacement measurements, and have many applications.  Our aim was to
use squeezing to achieve simultaneous squeezing of multiple
displacement measurements.  One of the primary experimental challenges
was to develop methods to spatially overlap two beams with orthogonal
spatial modes efficiently.  We achieved this overlap with approximately
95~\% efficiency using a specially designed optical resonator.  We used
this resonator to spatially overlap a pair of quadrature squeezed beams
with specifically configured mode-shapes.  Combining the resulting beam
with a bright coherent beam, also with a specific mode-shape, produced a
2D spatially squeezed beam.  This was demonstrated by simultaneous
below shot noise fluctuations in both horizontal and vertical
displacement measurements when the beam was imaged onto our quadrant
detector.  We introduced a spatial modulation to the beam and
demonstrated that, using spatial squeezing, a modulation amplitude of
1.6~\AA~could be detected with 99~\% confidence, a coherent beam, on
the other hand, could only detect modulations with amplitude greater
than 2.3~\AA. To illustrate the generality of out results, we also
demonstrated that horizontal and diagonal displacement measurements
could be simultaneously enhanced.

\section{Acknowledgment}
We acknowledge financial support from the Australian Research Council
Centre of Excellence Program and the European network QUANTIM, contract
IST-2000-26019.  We thank Roman Schnabel, Ben Buchler and Ulrik
Andersen for assistance and fruitful discussions.  The Laboratoire
Kastler-Brossel of the Ecole Normale Superieure and University Pierr{\' e}
et Marie Curie are associated via CNRS.


\end{document}